\documentclass[sigconf,nonacm,10pt]{acmart}

\usepackage{booktabs} 

\graphicspath{{figure/}{figures/}}

\usepackage{graphicx}
\usepackage{subcaption}
\usepackage{multirow}
\usepackage{tabularx}

\begin{document}
\title{Responsible developments and networking research: a reflection beyond a paper ethical statement}

\author{D. Tuncer}
\affiliation{%
  \institution{Ecole des Ponts ParisTech}
  \country(France)
}

\author{M. Bruyere}
\affiliation{%
  \institution{IIJ Research Lab and JFLI, CNRS}
  \country(Japan)
}

\begin{abstract}
Several recent initiatives have proposed new directions for research practices and their operations in the computer science community, from updated codes of conduct that clarify the use of AI-assisted tools to the inclusion of ethical statements and the organization of working groups on the environmental footprint of digitalization. In this position paper, we focus on the specific case of networking research. We reflect on the technical realization of the community and its incidence beyond techno-centric contributions. In particular, we structure the discussion around two frameworks that were recently developed in different contexts to describe the sense of engagement and responsibilities to which the practitioner of a computing-related area may be confronted.  
\end{abstract}

\maketitle
\pagestyle{empty}

\emph{The views expressed in this paper are the authors’ own and do not represent positions or opinions of their institutions. No automated text / image generation tools were used to produce this paper.}

\section{Introduction}
\label{sect:Intro}

In a 2011 position paper \cite{rexford11}, Jennifer Rexford shared with the networking community not only her enthusiasm with contributing to a relatively new field of research, \textit{i.e.,} networking, but also her apprehension when asked to state in simple and yet comprehensive terms what networking research \textit{really} entails. Fifteen years have almost passed since Rexford's attempt to "crystallize" the essence of the networking domain\footnote{As per indicated in the paper, the content is based on the talk given at the ACM SIGCOMM CoNext'10 Student workshop}. In the course of the last fifteen years, the on-going digital transformation of economies and societies in most parts of the world \cite{unesco21}\cite{worldbank23} has been infusing dynamics - many of them with worldwide scope - that are contributing to the future orientation of the Anthropocene\footnote{See \cite{pavid23} for a definition.}. While the process of digitalization kicked off far back in the past\footnote{While dating the exact beginnings of digitalization can be a daunting task, they are often associated with the work of Claude Shannon on the theory of information in the second half of the previous century, and in particular through his seminal paper on 'The Mathematical Theory of Communication' \cite{shannon48}.}, the realization of a world driven by digital developments reached a new step at the dawn of the twenty first century where the concurrence of rapid technological developments in multiple disciplines of computer science and engineering has enabled the shaping of a world of ubiquitous, speed-of-light connectivity. 

The digital shift of the connected world has been accompanied by a profound mutation of the characteristics of the interactions that humans and their activities induce to their environment, both at the physical (global footprint) and temporal (acceleration) scales. The effects of these mutated interactions are to be apprehended in a general context of events of strong interconnected nature that go far beyond the local and uni-dimensional scale \cite{zscheischler18}\cite{rgs22}, \textit{i.e.,} climate change, environmental crisis, political instability and conflicts, pandemics. In fact, while many of these events (and their impacts) were long described, discussed and even anticipated, they appear to manifest today in scales that make them more and more tangible, especially from outside restricted groups of concerned / alerted individuals \cite{nytimes131122}. In the recent years, several voices have been calling for the development of joint transnational efforts that directly address them as a whole \cite{nytimes131122}\cite{ramos20}.

Research and Innovation programs have largely been contributing to fostering the digital transformation, in particular through massive public and private investments in the development of automated, connected and data technologies. For instance, the current Digital Europe Program that covers the 2021 to 2027 period comes with a budget of 7.588 billion euro to support and accelerate the digital transformation of the "European economy, industry and society". Scientific and technological advances bring forward new paths for building knowledge and hence participate in defining the future of the digital world. The multi-scalar implications of digitalization, especially in the face of interconnected events, do however question the way today's research and its communities function. In a world where one's omniscience has long become impossible\footnote{Something that has been emphasized since 1945 - see Vannevar Bush \cite{bush45}}, it seems that it is becoming more than necessary to pause and think about the role and effectiveness of research (and its organization) in contributing to the future.

A wave of initiatives has recently been engaged in the various communities of computer science and engineering to (re)think / revisit the practices of a research that is an integral part of world's developments. Beyond the widely reported (and necessary) debates on the ethical concerns that are raised by the rapid development of AI (and its moral implications at every level of human organizations) \textit{e.g.,} \cite{deepMind21}, these initiatives touch on the different aspects of the traditional model of research, from updated codes of conduct (\textit{e.g.,} ACM Code of Ethics and Professional Conduct \cite{acmEthics}, IEEE Global Initiative for Ethical Considerations in Artificial Intelligence and Autonomous Systems \cite{ieeeEthics}), discussed topics (\textit{e.g.,} ACM SIGCOMM 2022 Joint Workshops on "Technologies, Applications, and Uses of a Responsible Internet" and "Building Greener Internet" (TAURIN + BGI) \cite{TAURIN22}, social responsibility in game AI \cite{cook21}), research assessment protocols (\textit{e.g.,} inclusion of required ethical statement for reviewers and paper's authors) or training programs for doctoral students and researchers (\textit{e.g.,} development of ethical reasoning skills for computer scientists \cite{grosz19}\cite{harvard}, online MOOCs on Ethics and ICT \cite{moocPSU23}). These initiatives all participate in evolving, in the various areas of computer science and engineering, what research is about in the twenty first century, how it is conducted and what role researchers play with respect to the society and the environment. 

This position paper follows up on these initiatives by providing an inevitable personal perspective on these questions through the reference to two frameworks recently developed as a kick-off to the thinking process about ethics and digital developments, namely the work of the pilot committee for digital ethics started in France in 2019 \cite{CNPEN22} and the four-level scale model of ethical engagement proposed by Chiodo and Bursill-Hall in the UK in the domain of mathematics \cite{chiodo18}. The objective is to open the debate among the computer network research community and introduce these questions as core components of the research carried out. In particular, we use here the framework introduced by Chiodo and Bursill-Hall \cite{chiodo18} to discuss how researchers in the community can position with respect to the moral and ethical implications of their work contributions, ranging from finding them irrelevant to feeling highly concerned and wishing to invest time in addressing them. As noted by Chiodo and Bursill-Hall, this categorization is to prone to be subjective to one's background, history and identity, and as such, is not to be taken as an absolute (or even worst influencing) guiding principle for setting goals. The exercise enables us however to structure a discussion on questions we believe are important to be dealt with as and within a community. We are concerned\footnote{as in having a sense of responsibility} by the impact of our research developments, recognize that our concerns are shared with others, and wish these questions to become non-taboo in the domain to which we have been contributing.

The paper is organized around four core sections. Section \ref{sect:netwResearch} provides an overview of research in networking. Section \ref{sect:beyondTech} introduces a perspective on the implications of networking research developments beyond the technical domain. Section \ref{sect:responsibility} discusses the sense of responsibility and ethical engagement through the two frameworks. Section \ref{sect:nextsteps} presents a set of possible next steps on how these questions can be concretely tackled and be part of the scientific debate in our community. Concluding remarks are presented in Section \ref{sect:conclusion}.
 
\section{Networking research today} 
\label{sect:netwResearch}

The story of networking follows the story of human communications. Humans have been employing various communication strategies to exchange data since the mist of time. While we relied on direct natural resources as support for communication for most of our past history (\textit{e.g.,} information exchange through horse carriers), a significant shift in the comprehension of electromagnetism in the nineteen century paved the way to our modern-day communication systems. From analog to digital communications, today's networks are the results of progress in telecommunication, computer, electronics and material science technologies.   

In today's networking research, a network is the object that represents the (inter)connections between sets of heterogeneous components that exchange data, from devices and machines, to systems. The connection can be physical, \textit{i.e.,} component A and component B are connected through a physical link, but it can also be logical, \textit{i.e.,} there is a relationship between component A and component B. The data to exchange is \textit{transported} through an infrastructure that comprises various hardware and software components using different technologies such as optical fibers, antennas or satellite. Research in networking can thus be broadly described as investigating how to achieve these exchange of data, including to study what infrastructure is needed, to analyze how to orchestrate the communication, to understand the structure of these connections, or as stated by Schulzrinne \cite{schulzrinne18}, to improve a "core civilisational infrastructure". 

\begin{figure*}[ht]
  \centering
  \includegraphics[width=\linewidth]{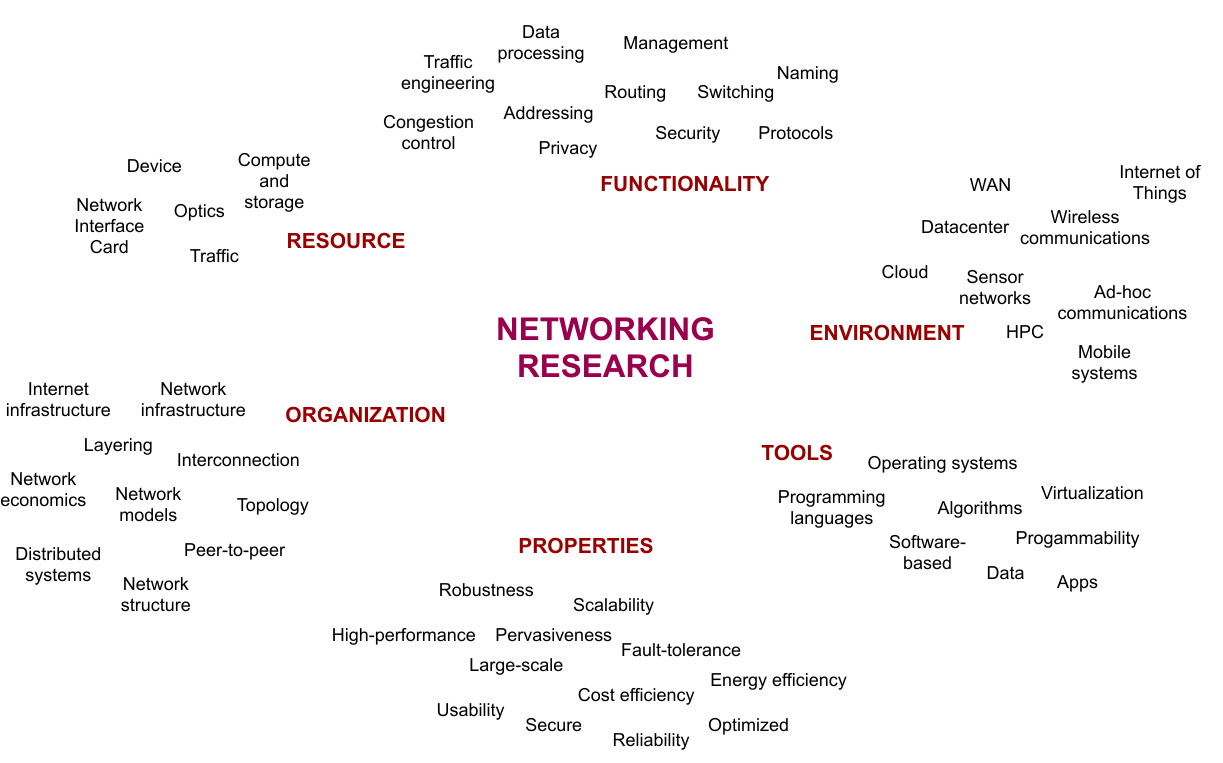}
  \caption{Landscape of networking research used expressions and keywords from the researcher themselves.}
  \label{fig:landscape}
\end{figure*}

\subsection{Networking research in practice}
\label{sect:netwPract}

This broad definition of networking research partially reflects the diversity of problems to which researchers in networking are exposed. As emphasized by Jennifer Rexford \cite{rexford11}, the problems tackled by researchers in networking tend to fall under and across traditional disciplines, which makes it difficult to draw an exhaustive and precise picture of what research in this domain actually is. In fact, it may even be dubious to try and elaborate a definite strictly-framed definition of networking research. A more realistic and attainable objective is to try and provide a perspective on its diversity. We suggest three complementary approaches to survey the landscape of networking research:
\begin{itemize}
	\item \textbf{Approach 1} to review the traditional nomenclature used by the main relevant professional organizations, such as ACM\footnote{\url{https://dl.acm.org/ccs\#}} or IEEE\footnote{\url{https://www.ieee.org/publications/services/thesaurus-thank-you.html}}, to categorize publications.
	\item \textbf{Approach 2} to do a systematic keyword/text analysis on the calls of papers of events that fall under the scope of "networking" research. 
	\item \textbf{Approach 3} to directly probe the researchers who identify themselves as "researchers in networking" to \textit{listen} to what they have to say about the research they do. 
\end{itemize}

Each approach has pros and cons. Approach 1 is normative but restricts the definition of the research problem space to a pre-defined fixed set of categories. Approach 2 captures the evolution of interests but assumes a pre-definition of networking research in order to identify relevant venues. Approach 3 is focused on the researchers and is useful to get perceptions but is biased by the ability to describe one's work. While the ideal strategy to effectively draw the landscape of networking research would be to combine the three approaches, we focus here on the third one \footnote{The objective is here to discuss our practices, not to propose a definition of networking research.}.

\subsection{Listening to researchers}
\label{sect:listening}

We propose to "listen" to the researchers by consulting the web pages describing the activities of their research group. We believe that this is a good proxy for listening given that web pages are likely to be written by the researchers themselves. While it is unavoidable that directly focusing on the words of researchers introduces a level of uncontrolled subjectivity, it is likely to give a better sense of how researchers perceive what their research is about. To do so, we manually consulted the web pages of a set of academic research groups that host members who had a paper published in one of the four following networking-classified venues, namely SIGCOMM, NSDI, CNSM and IMC. The venues were selected based on their reputation to represent different domains traditionally associated with networking research, including general computer networks (SIGCOMM), network systems (NSDI), network management (CNSM) and network measurements (IMC). We consulted the web pages of a total of $18$ research groups\footnote{The list is provided in Appendix \ref{apdx:researchGroup}.}. All research groups are hosted by academic institutions that are located in Asia, North America, Brazil or Europe\footnote{It is (unfortunately) worth noting the absence of any publications from authors affiliated with institutions in Africa, other South American countries, or Oceania in the considered venues over the considered time period}. On each web page, we surveyed the type of research topics reported by the group, as well as the keywords used to describe the work carried out by the group members. 

Figure \ref{fig:landscape} shows the main keywords and expressions used on the web pages of the surveyed research groups that we categorized in six general themes, namely Organization, Resource, Functionality, Environment, Tools and Properties. They all refer to a range of technical specifics of networks, whether in terms of equipment, functions or infrastructures. More specifically, research topics dealing with \textbf{Organization} aspects include all research that has to do with studying the arrangement of networks, either as systems, through their infrastructure, their connectivity models or their institutional groupings. Topics pertaining to \textbf{Resource} involves research efforts that investigate the resources needed for the network to exist. Topics on the \textbf{Functionality} includes the range of capabilities provided by a network, \textit{e.g.,} from conveying data from point A to B to ensuring that the data is delivered in a secured fashion. Topics related to the \textbf{Environment} concern all efforts that focus on studying networks in different contexts, from performing high performance computing tasks, to enabling exchange of information between sensors for instance. Topics focusing on \textbf{Tools} investigate the tools required to manipulate and handle data in a digital form for distribution and exchange across machines. Finally, research around the \textbf{Properties} of networks represent all efforts that look into the characteristics of sending, transporting and delivering digital data. 

The figure shows the diversity of topics covered by networking research. While the list is surely not exhaustive, it illustrates well that the primary focus of networking research is technical. The work is generally hosted under Faculty / School of Engineering and the approaches typically employed are hard science-based, \textit{e.g.,} quantitative analysis, experimentation, modeling. The questions addressed in the literature spread over different flavors, from being engineering-focused (optimization to make system work efficiently) to system-based (building systems to test their feasibility) or physics-inspired (understanding the underlying principles and rules that make networked systems the way they are). While networking research has mainly been focused on the technicality of computer systems and their operations, the implications of its developments do however far exceed the sole technical system perimeter. 

\section{Beyond technical dimensions}
\label{sect:beyondTech}

As a domain of applied research, networking embeds the "societal value found in improving a core civilizational infrastructure" \cite{schulzrinne18}. In particular, it is hard to conceive, in a digital world, computer networks and their evolution as passive agents that do not "exploit and influence social or physical phenomena" and "change the world through their use" \cite{chiodo23}. A step aside the technical focus of typical networking research is thus a good exercise to embrace the domain from a different angle and reason in terms of the "social and physical" axes of the research developments. We propose to organize these axes around five main dimensions, namely Time, Space, Interaction, Automation and Environment. 

\subsection{Networking research and time} Time is a key dimension of networking research. The evolution of computer networks has made possible the delivery of very diverse types of content that only gets constrained by the physical capabilities of the underlying technology (\textit{e.g.,} optics, radio). Time-related metrics are central to the evaluation of performance of a networking solution, \textit{e.g.,} whether in terms of reduced latency or improved speed. 

Time is a hard notion in science and in fact, the existence of time when factoring in the level of observations is not even a given \cite{livni18}. From a societal perspective, however, networking research has been accompanying a fundamental shift towards a perceived sense of acceleration \cite{gomez18}, something well-illustrated by some of the currently hot topics in networking such as ultra-low latency and high-speed networking. These come with a vision of a future\footnote{Whether desirable is another question.} where the network acts as the medium of immediate proximity through the applications it supports, \textit{e.g.,} holographic communication, augmented live experience.

At the same time, the divide between regions in terms of infrastructure coverage strongly challenges the idea of a global access to \textit{any} type of content and by extension (see for instance the recent work by Habib \textit{et al.} \cite{habib23}) the provision of uniformly distributed digital services. From a time perspective, this raises questions with respect to the way networking research contributes to shaping time. It is not simple to determine whether networking research is working towards creating global, temporal uniformity, \textit{i.e.,} a model of instantaneous delivery, or is in contrast accentuating the divergence in terms of temporal perception across regions.    

\subsection{Networking research and space} Space is an inherent dimension of networking. The (still) active area of networking research on content- / information-centric networking \cite{jacobson07}\cite{chai11} is for instance an invitation to rethink the spacial dimension in the context of networks. More specifically, space in networking can be in terms of the physical space, \textit{i.e.,} an infrastructure for different geographical settings (\textit{e.g.,} personal -, local -, wide area network). Research onto edge cloud for instance builds on the embedded notion of space when it states that it brings content closer to the users. The space can also be in terms of geopolitical space. Research efforts on Schengen routing constitute illustrative examples \cite{donni15}\cite{ochs16}, so do initiatives that investigate issues related to the (de)centralization of the Internet, \textit{e.g.,} \cite{raman19}\cite{harchol20}. Finally, the space is also a matter of administrative boundaries. Networks are by nature distributed infrastructures that follow an organization in domains that can cross these boundaries.  

The outcomes of networking research contribute to the development of digital services that question our relation to space, \textit{e.g.,} e-commerce and urban logistics, virtual location viewing, tele-medicine. While the network functionality (\textit{i.e.,} to deliver bits from point A to point B) is supposed, by design, to be agnostic to these applications, its realization necessitates physical resources that have a physical footprint and a spacial existence. All these directly interrogate the spacial reality of digital solutions and their supporting networks.    
 
\subsection{Networking research and interaction} A computer network is a medium of interactions. It enables communication between agents, the agents being either the direct recipient of the data (\textit{i.e.,} the machine) or the indirect target, \textit{i.e.,} the people. The effects of networking research developments on the time and space dimensions, respectively, manifest in a combined manner on interactions. Networking technologies have come in support of a model of ubiquitous connectivity that constitutes the starting point of a major digital shift in many sectors, \textit{e.g.,} public services, education, banking, industry. These changes bring new modes of organization and new business models, which induces significant societal implications\footnote{See for instance the documented case of France \cite{jeannot22}.}. These new models themselves influence the orientation of research developments in terms of strategic topics and funding. They also have a direct incidence on the purpose of the undertaken research. The outcomes of a research on community network \cite{fuchs17}, for instance, that works for the common good are surely in antinomy to the outcomes of a research that looks into the management of mega-scale data center infrastructures for commercial deep-learning-based applications. The forms of interactions enabled by networks are thus strongly driven by the socio-economic models that these networks have enabled to shape. 

\subsection{Networking research and automation} The vision of fully autonomous networks capable of self-configuring, self-healing, self-maintaining, has been a core objective to achieve in the networking community for a long time. Over time, this vision has been expressed in different ways, usually in line with buzzing computer science interests of the time, \textit{e.g.,} autonomous networking and homeostasis from bio-inspired computing, self-driven networks as an analogy to self-driving vehicles, zero-touch networking and its AI-based networking successor, and \textit{tutti quanti}. 

Despite rebranding effects, the challenges raised by the realization of networks that can operate by themselves fundamentally remain unchanged with various support from the evolving technology, \textit{e.g.,} software-defined networking, programmability, virtualization. In this quest to a zero-touch system, it is surprising to notice that understanding the people who make the networks work has been to a large extent understudied. In the literature, human intervention is frequently referred to as synonymous to sources of errors, and the motivations advocated for less and less human involvement assume a direct impact on a higher degree of performance, a model by which developments are purely metric-driven. The people behind the operations of networks have accumulated experiences that make their know-how invaluable to understand the \textit{actual} needs of automation and make it happen in a manner that is human-focused rather than quantitative performance-driven, a long way from current practices. 

\subsection{Networking research and environment} In December 2022, the Internet Architecture Board organized a workshop on the environmental impact of Internet applications and systems \cite{iab22} with the objective to "bring together a diverse stakeholder community to discuss" the environmental impact of the Internet and "evolving needs from industry" and "areas for improvements and future work". Topics related to energy efficient networking are not new in the community (see \cite{bolla10}\cite{bianzino10}\cite{ruth09} for instance). Their scopes are however today defined in a larger dynamic of research that investigates and evaluates the environmental footprint of human activities, with digitalization still an under-covered contributor \cite{schwarzer23}\cite{shift23}. Networking developments are not exempt to rebound effects \cite{schneider01}, the today's well-documented phenomenon by which facilitating the access and the use of a technology favors an increase in consumption. Recognizing these impacts, developing methods to describe and evaluate them (\textit{e.g.,} \cite{clemm23}), and engaging the community as a whole to addressing these objectives seem a fundamental action. \\

The five dimensions discussed in this section constitute neither independent nor isolated blocks, and in general, they pose questions that necessitate a joint problem-solving approach. While networking research is certainly relevant to investigating these questions, its impact is likely to be limited if not combined with perspectives from other disciplines and domains. Today's practice in the community makes it difficult to adopt a multi-domain position when reporting / presenting results and methods. Despite the incentives towards embarking multiple disciplines in projects, outputs are often too focused on what each community expects to see. In that respect, networking having developed as a research domain that aggregates mixed practices can be pivotal to test new formats for conducting and reporting research.

\section{Sense of engagement and responsibility} 
\label{sect:responsibility}

To paraphrase Toyama and Ali \cite{toyama09}, it is legitimate to ask whether the issues discussed in the previous section are relevant to networking research. Drawing lines between what is or is not networking research is however problematic, especially for a domain that is well-known to cross disciplines (computer science, electronic and electrical engineering, applied mathematics). The answer to this question is likely to depend on one's perception of their sense of responsibilities with respect to what they do, a debate that is far from being new in science, \textit{e.g.,} see the examples of Wiener \cite{wiener49}, Grothendieck \cite{grothendieck72} or Weizenbaum \cite{weizenbaum76}. In this section, we propose to take a look at two recent frameworks that try to describe more formally the sense of engagement and responsibilities to which the practitioner of a computing-related area may be confronted.  

\subsection{Grand questions} 
\label{sect:cnpen}

In the face of rapid developments in digital and AI technologies, a pilot national committee for digital ethics\footnote{Comit\'e consultatif national d'\'ethique - \url{https://www.ccne-ethique.fr/en/cnpen}} was established in December 2019 in France. Its role is to discuss, assess and provide information and support relative to the ethical challenges raised by these technologies that individuals, the civil society, public and private institutions, as well as the government, have to face today. The committee involved 27 members from different disciplines, including computer science, law, philosophy, history, \textit{etc.} and sectors, \textit{i.e.,} academia, civil society, public administration, industry and innovation. The founding motive for the work undertaken was highly inspired by the crucial role of ethics committees in the area of health and medical science. 

\begin{table*}[t]
\caption{Networking research topics and ethical challenges.}
\label{table:topicsandethics}
\renewcommand{\arraystretch}{1.2} 
\renewcommand{\tabcolsep}{0.175em}
\footnotesize
\centering
\begin{tabular}{|c|c|}
\hline
\textbf{Grand question} & \textbf{Example relevant topics} \\
\hline
Redefinition of the relationship to ourselves and to others & Privacy, social networks\\
\hline
Autonomy and fight against the digital divide & Internet accessibility, 5/6G\\
\hline
Sovereignty in democratic choices & Centralized \textit{vs.} decentralized control\\
\hline
Free will in relation to human beings and machines & Network automation\\
\hline
Sustainability of digital developments & High-speed networks, hyper-scale infrastructure\\
\hline
\end{tabular}
\vspace{-1mm}
\end{table*}

In 2022 the committee released a report \cite{CNPEN22}, infused by many practical use case studies, that highlights a series of grand questions that invite one's to think about the ethical impact of developments in the general domain of computer science, more specifically with respect to:  
\begin{itemize}
    \item The redefinition of the relationship to ourselves and to others, in all spheres of social relations: family, friends, work and institutions.
    \item The means to respect the autonomy of people and to fight against the digital gap.
    \item The ability to remain sovereign in our democratic choices.
    \item The conditions to exercise our free will in a relationship between human beings and machines.
    \item The social, economic and environmental sustainability of digital developments.
\end{itemize} 

These challenges, on which traditional topics of interest in networking, such as privacy, centralized control or hyper-scale infrastructure, have direct implications, offer a relevant reading grid for networking research developments. Table \ref{table:topicsandethics} provides some examples of topics related to networks. Such a framework could be used for instance to revisit and adapt the nomenclature and presentation of the topics of interest listed on the web pages of the different venues of networking research. It could also inspire the organization of workshops and tutorials, hosted in general networking events, that accept contributions from a larger community. 

\subsection{Ethical engagement in levels}
\label{sect:framework}

From their practice in mathematics, Chiodo, Bursill-Hall and Mueller observed that researchers tend to be grouped into different categories with respect to the time they wish to invest in thinking about their responsibilities towards the developments to which they contribute. In particular, they propose a four-level scale \cite{chiodo18} to describe these different forms of ethical engagement, \textit{i.e.,} absence of any sense of responsibility (the lowest level - level 0), awareness of ethical challenges (level 1), readiness for actions (level 2) and active advocacy for engagement (highest level - level 3). 

The availability of a gradation-based framework based on qualitative rather than quantitative attributes is highly effective for reasoning upon the possible ethical implications and paths a development can take. The specific case of the IP header design is a good example in networking to illustrate how thinking in terms of time invested for engagement can help structure a debate on the consequences of a design decision. The format of IP packets was designed in the seventies. At that time, designers did not spend time thinking about the large scale latency impact of processing packet fields and as such, the fact of inserting the source address before the destination address in the packet header did not seem to be problematic. As a result, routers today need to wait a processing time of 4 bytes before knowing where a packet needs to be routed, which, multiplied by the number of routing decisions taken in an operational context, can easily involve processing costs that are not insignificant. The issue may appear as minor for some (level 0) while others may recognize the implications of this design choice (level 1) and may even invest time to evaluate its impact in terms of latency and power consumption (level 2). Yet others may want to go a step even further in order to devote time and be in position to arbitrate on the required aspects to take into account (including the environmental footprint) before a protocol design choice can be validated (level 3).  

While initially developed for research in mathematics, the proposed scale-based framework builds upon general observations that can easily be transposed to the networking research context. The authors of the framework recently developed a series of exercises \cite{chiodoTeaching} that can be used to raise awareness about the ethical implications of the mathematics object. This comes as a set of illustrative problems, each associated with a set of questions that cater for various degrees of investment in ethical thinking. Developing a similar handout for the network object can provide a useful tool for the community to work on and embed ethics in project developments. This necessitates however to identify a set of representative and yet easy to describe networking problems for which levels of ethical considerations can be formulated. We are currently working on this.      

\section{Possible next steps} 
\label{sect:nextsteps}

Questioning the role of techniques and technologies, as well as their constructs, is the focus of research in sociotechnical systems \cite{cooper71}. The primary purpose of networking research is technical developments and to continue investigating these developments is fundamental. It is however problematic and questionable to pursue them in a mode that addresses problems from their technical perspective only. While the role of networking research is not to elaborate theories around the purpose of techniques and technologies, it can benefit a lot from stepping aside its pure technicality and regard problems through their multidimensional nature. We discuss here three general concrete actions to contribute to this objective.  

\textbf{A shared space for ideas} Workshops can be a good setting to share ideas and debate on topics that go beyond the technical perspective of networking developments. This format has been attempting in different venues, in particular at SIGCOMM (ACM SIGCOMM 2022 Joint Workshops on "Technologies, Applications, and Uses of a Responsible Internet" and "Building Greener Internet", ACM SIGCOMM 2023 Second Workshop on Situating Network Infrastructure with People, Practices, and Beyond). We believe that venues should encourage the submission of workshop proposals that are less formal than traditional workshops that are usually structured around published papers, authors' talk and proceedings. These workshops should invite participants to think out of the box and only encourage the submission of fully polished papers on a volunteering basis, and not as a prerequisite to being secured a time slot. In addition, the format should encourage active participation, similar to flipped classroom models, where attendees work together on the co-construction \cite{jacoby95} of propositions, questions and / or suggestions. In international venues, this would be an opportunity to get the perspectives of different parts of the world, and challenge mainstream positions.   

\textbf{Beyond ethical statement as a tick box exercise} Many networking conferences started to introduce the requirement of an ethical statement for submitted and published papers. While this initiative can only be positively welcome, it is a concern that without clear guidance on what an ethical statement entails, the process could end up in a box ticking exercise. The four-level scale framework \cite{chiodo18} shows well that one's perception of the sense of responsibility is highly subjective and variable between people. As a concrete output of the initiative they started eight years ago, Chiodo and Mueller developed a very detailed methodology, that they present in the form of a manifesto \cite{chiodo23}, for identifying and addressing every aspect of ethical questioning when working on and designing a solution to a mathematical problem. The methodology involves ten main domains, that the authors refer to as pillars, each divided in a series of detailed questions and sub-questions to take into account. We believe that an approach inspired by this methodology and applied to the context of networking research would be very valuable for enhancing the current practice of filling in ethical statements.

\textbf{The choice of naming} There have been for a long time animated debates around the relevance of the world "intelligence" when talking about the ability of computing machines \cite{searle80}\cite{lani23}. The hype around expressions and wordings can have an influence on the intellectual construction of ideas (see \cite{alpuim23}). The choice of words is not neutral, it embeds deeply ingrained beliefs (see research in linguistics, \textit{e.g.,} \cite{shashkevich19}). Yet networking research venues do not typically host sessions focusing on the epistemology of its development. In a research that is highly international, with contributions from a lot of non-native English speakers, it does not seem absurd to open venues to new forms of activities that focus on investigating and reviewing the choice of naming in the community. Networking researchers are surely in the best position to share a perspective on these questions.

\section{Concluding remarks}
\label{sect:conclusion}

We are certainly not the first to question the implication of our research contributions. The rapid evolution of the Internet infrastructure over the past few decades has transformed the landscape of our global human society. With approximately 65\% of the world's population now connected online, the pervasive nature of digitalization is but evident. While widespread connectivity has brought many opportunities and benefits, it has also given rise to significant challenges. Today it has become essential to address research questions through which the Internet and computer networking technologies are envisioned to be aligned with human- and environment-centric values. Computer networking researchers cannot ignore or avoid ethical questions, and we believe that developing and implementing new activities as part of the construction, communication and sharing of knowledge is a promising, concrete next step.

\appendix

\section{Surveyed research group webpages}
\label{apdx:researchGroup}

\begin{table*}[t]

\caption{List of consulted research groups.}
\label{table:researchGroup}

\renewcommand{\arraystretch}{1.2} 
\renewcommand{\tabcolsep}{0.175em}
\footnotesize
\centering
\begin{tabular}{p{4.6cm}p{4.6cm}p{2.2cm}p{4.6cm}}
\hline
\textbf{Institution} & \textbf{Group / Area} & \textbf{Country} & \textbf{URL}\\
\hline
Tsinghua University & Computer Networks & China & \url{https://www.cs.tsinghua.edu.cn/csen/Research/Research_Areas/Computer_Networks.htm}\\
\hline
Hong Kong University of Science and Technology & Networking and Computer System & China & \url{https://cse.hkust.edu.hk/pg/research/themes/}\\
\hline
The Hong Kong Polytechnic University & Networking and Mobile Computing & China & \url{https://www.polyu.edu.hk/comp/research/research-groups/} \\
\hline
POSTECH & Network Systems & Korea & \url{https://ecse.postech.ac.kr/research-activities/research-field/} \\
\hline
Princeton University & Systems \& Networking & USA & \url{https://www.cs.princeton.edu/research/areas/systems} \\
\hline
University of Washington & Computer Systems \& Networking & USA & \url{https://www.cs.washington.edu/research/systems} \\
\hline
University of California-San Diego &  Systems and Networking & USA & \url{http://www.sysnet.ucsd.edu/sysnet/} \\
\hline
Standford University & Software systems, security, distributed systems and networks & USA & \url{https://ee.stanford.edu/research/software-systems} \\
\hline
Massachusetts Institute of Technology & Networks at MIT & USA & \url{https://www.csail.mit.edu/research/networks-mit} \\
\hline
Cornwell University & Systems and Networking & USA & \url{https://www.cs.cornell.edu/research/systems} \\
\hline
University of California, Berkley & Operating Systems \& Networking & USA & \url{https://www2.eecs.berkeley.edu/Research/Areas/OSNT/}\\
\hline
Technical University of Munich & Computer \& Communication Systems & Germany & \url{https://www.ce.cit.tum.de/en/ce/research/areas/communications/} \\
\hline
Max Plank Institute & Network and Cloud Systems & Germany & \url{https://www.mpi-inf.mpg.de/departments/network-and-cloud-systems} \\
\hline
University Catholique de Louvain & Communication Systems and Networks & Belgium & \url{https://uclouvain.be/en/research-institutes/icteam/communication-systems-and-networks.html} \\
\hline
ETH Zurich & Computer Systems & Switzerland & \url{https://inf.ethz.ch/research/computer-systems.html} \\
\hline
University College London & Systems and Networks & UK & \url{https://www.ucl.ac.uk/computer-science/research/research-groups/systems-and-networks-research-group}\\
\hline
University of Cambridge & Systems and Networking & UK & \url{https://www.cst.cam.ac.uk/research/themes/systems-and-networking} \\
\hline
Federal University of Rio Grande do Sul & Computer Networks & Brazil & \url{https://www.inf.ufrgs.br/site/en/research/research-groups/computer-networks/} \\
\hline
\end{tabular}
\vspace{-1mm}
\end{table*}

\end{document}